\documentclass[a4paper]{jpconf}

\usepackage{bm}
\usepackage{amsmath,amssymb}
\usepackage[usenames]{color}

\usepackage{graphicx}

\def\journal #1#2#3#4{#4 {\it #1} {\bf #2} #3}

\def\PRB{Phys.\ Rev.\ {\rm B}}
\def\PRL{Phys.\ Rev.\ Lett.}

\def\JPSJ{J.\ Phys.\ Soc.\ Jpn.}

\begin{document}
\title{Low-energy Majorana states in spin-liquid transitions in a three-dimensional Kitaev model}

\author{Joji Nasu$^{1}$, Masafumi Udagawa$^{2}$, and Yukitoshi Motome$^{2}$}

\address{$^{1}$Department of Physics, Tokyo Institute of Technology, Ookayama, 2-12-1, Meguro, Tokyo 152-8551, Japan}
\address{$^{2}$Department of Applied Physics, University of Tokyo, Hongo, 7-3-1, Bunkyo, Tokyo 113-8656, Japan}

\ead{nasu@phys.titech.ac.jp}

\begin{abstract}
A three-dimensional Kitaev model on a hyperhoneycomb lattice is investigated numerically at finite temperature. 
The Kitaev model is one of the solvable quantum spin models, where the ground state is given by gapped and gapless spin liquids, depending on the anisotropy of the interactions. This model can be rewritten as a free Majorana fermion system coupled with $Z_2$ variables. 
The density of states of Majorana fermions shows an excitation gap in the gapped region, while it is semimetallic in the gapless region reflecting the Dirac node. Performing the Monte Carlo simulation, we calculate the temperature dependence of the Majorana spectra.
We find that the semimetallic dip is filled up as temperature increases in the gapless region, but surprisingly, the spectrum develops an excitation gap in the region near the gapless-gapped boundary.
Such changes of the low-energy spectrum appear sharply at the transition temperature from the spin liquid to the paramagnetic state. 
The results indicate that thermal fluctuations of the $Z_2$ fields significantly influence the low-energy state of Majorana fermions, especially in the spin liquid formation.
\end{abstract}

\section{Introduction}

Quantum spin liquid (QSL) is one of the fascinating subjects in condensed matter physics~\cite{Balents2010}. 
This is a new state of matter in magnetic insulators, which does not show long-range magnetic ordering down to zero temperature ($T$). 
Vast experimental efforts have been made to realize this exotic state, and several candidates of QSLs were proposed, for instance, in organic salts~\cite{Shimizu03,Yamashita10} and transition-metal oxides~\cite{Nakatsuji05,Helton07,Okamoto07}.
Theoretical studies have also been performed for many models, e.g., the Heisenberg and Hubbard models on geometrically frustrated lattices~\cite{Morita2002,Yan2011,Jiang2012}.
In spite of these intensive studies, it remains controversial whether QSLs are realized or not in the theoretical models, mainly because of the difficulty in numerical simulations, such as the negative sign problem in the quantum Monte Carlo (MC) method.

The Kitaev model is a quantum spin model consisting of $S=1/2$ spins~\cite{Kitaev06}.
This model is originally defined on a two-dimensional (2D) honeycomb lattice composed of three types of bonds.
On each bond, the interaction between the nearest-neighbor spins is of Ising type but the spin component of the interaction is different among the three types of bonds. 
This bond-dependent interaction brings about frustration; namely, all the bond energies are not minimized simultaneously. Due to the frustration effect, a magnetic order is suppressed down to zero $T$ and a nontrivial magnetic state emerges in the ground state. Indeed, the ground state of the Kitaev model is exactly proved to be a QSL~\cite{Baskaran2007}.
Depending on the anisotropy of the exchange interactions, both gapped and gapless QSLs appear in the ground state~\cite{Kitaev06,Baskaran2007}.
Therefore, this model provides a good starting point to reveal the intrinsic properties of QSLs. 

In addition, it was proposed that the Kitaev model is relevant also experimentally: the Kitaev-type  interaction may be realized between $j_{\rm eff}=1/2$ spins under the strong spin-orbit coupling in iridium oxides with a layered honeycomb lattice~\cite{Jackeli2009}.
Recently, related new iridium compounds, in which the iridium ions form a three-dimensional (3D) network were synthesized in the chemical formula Li$_2$IrO$_3$: the so-called hyperhoneycomb~\cite{Takayama2014} and harmonic-honeycomb compounds~\cite{Modic2014}. In these 3D materials, the Kitaev-type interaction is also expected to be present.
The discoveries have stimulated theoretical studies of the 3D Kitaev physics~\cite{Lee2014,Kimchi1309,LeeSB2014,Nasu2014,Nasu2014p,Lee2014p,Kimchi2014p}.
Among them, the authors and their collaborators have clarified the existence of finite-$T$ phase transitions between the low-$T$ QSLs and the high-$T$ paramagnet by extensive numerical simulations~\cite{Nasu2014,Nasu2014p}.

In this paper, we address the finite-$T$ properties of the Kitaev model on a hyperhoneycomb lattice.
This 3D Kitaev model was first introduced in Ref.~\cite{Mandal2009}. 
One of the characteristics in the Kitaev model is that this model can be exactly solvable at zero $T$ by rewriting it as a free Majorana fermion system coupled with $Z_2$ variables. 
We here focus on the effect of thermal fluctuations of the $Z_2$ variables on the Majorana fermion state.
We calculate the $T$ dependence of the density of states (DOS) of Majorana fermions, and compare the results with the $T$ dependence of the specific heat of the total system. 
We find that the low-energy Majorana spectrum exhibits characteristic $T$ dependence around the transition temperature between QSL and paramagnet. 
We show that the gapless behavior is eroded by thermal fluctuations of the $Z_2$ fields above the critical temperature.

\section{Model}
\begin{figure}[t]
\begin{center}
\includegraphics[width=0.8\columnwidth,clip]{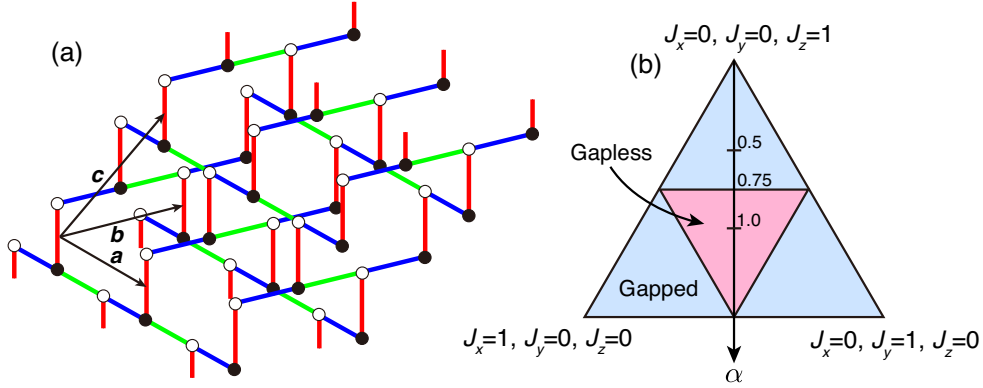}
\caption{
 (a) Lattice structure of a hyperhoneycomb lattice. The blue, green, and red bonds correspond to $x$, $y$, and $z$ bonds in Eq.~(\ref{eq:5}), respectively. $\bm{a}$, $\bm{b}$, and $\bm{c}$ are the primitive translation vectors. (b) Ground-state phase diagram of the 3D Kitaev model defined on the hyperhoneycomb lattice on the plane of $J_x+J_y+J_z=1$. There are two different QSL phases distinguished by the excitation.
The parameter $\alpha$ defined by $J_x=J_y=\alpha/3$ and $J_z=1-2\alpha/3$ is also indicated.
}
\label{fig_hyperhoneycomb}
\end{center}
\end{figure}

We study the Kitaev model on a hyperhoneycomb lattice, whose Hamiltonian is given by 
\begin{align}
{\cal H}=-J_x\sum_{\langle ij\rangle_x}\sigma_i^x\sigma_j^x-J_y\sum_{\langle ij\rangle_y}\sigma_i^y\sigma_j^y-J_z\sum_{\langle ij\rangle_z}\sigma_i^z\sigma_j^z,\label{eq:5}
\end{align}
where $\sigma_i^x$, $\sigma_i^y$, and $\sigma_i^z$ are Pauli matrices describing a spin-1/2 state at a site $i$;
$J_x$, $J_y$, and $J_z$ are exchange constants~\cite{Kitaev06}. 
The model is defined on the hyperhoneycomb lattice shown in Fig.~\ref{fig_hyperhoneycomb}(a)~\cite{Mandal2009}; 
the interactions $J_x$, $J_y$, and $J_z$ are defined on three different types of the nearest neighbor bonds, $x$ (blue), $y$ (green), and $z$ bonds (red), respectively.
The ground state of this model is obtained exactly, similarly in the 2D Kitaev model on a honeycomb lattice~\cite{Kitaev06}; the phase diagram is completely the same as that of the 2D model, and consists of gapless and gapped QSL phases, as shown in Fig.~\ref{fig_hyperhoneycomb}(b)~\cite{Kitaev06}.
The QSL with gapless excitation is stabilized in the center triangle including the isotropic case $J_x=J_y=J_z$, while the QSL with an excitation gap appears in the outer three triangles with anisotropic interactions.

\section{Method}\label{sec:method}
We study thermodynamic properties of the model in Eq.~(\ref{eq:5}) by an unbiased MC method. 
The method is based on the Majorana fermion representation of the model, described below. 
First, we apply the Jordan-Wigner transformation by considering the 3D hyperhoneycomb lattice as a set of one-dimensional chains composed of $x$ and $y$ bonds. 
These chains are connected by the $z$ bonds with each other. 
A site $i$ on the hyperhoneycomb lattice can be represented by a pair of integers $(m,n)$, where $m$ identifies a chain and $n$ is the site index on the $m$-th chain.
Then, by the Jordan-Wigner transformation, the spin operators are written by spinless fermion operators ($a_i$, $a_i^\dagger$) as
$S_{m,n}^+
=(S_{m,n}^-)^\dagger=\frac{1}{2}(\sigma_{m,n}^x+i \sigma_{m,n}^y)=\prod_{n'=1}^{n-1}(1-2n_{m,n'}) a_{m,n}^\dagger$
 and $\sigma_{m,n}^z=2n_{m,n}-1$,
where $n_i$ is the number operator defined by $n_i=a_{i}^\dagger a_{i}$.
The Ising-type interactions in Eq.~(\ref{eq:5}) are written as
$\sigma_{m,n}^x\sigma_{m,n+1}^x=-(a_{m,n}-a_{m,n}^\dagger)(a_{m,n+1}+a_{m,n+1}^\dagger)$, 
$\sigma_{m,n}^y\sigma_{m,n+1}^y=(a_{m,n}+a_{m,n}^\dagger)(a_{m,n+1}-a_{m,n+1}^\dagger)$, and 
$\sigma_{m,n}^z\sigma_{m',n'}^z=(2n_{m,n}-1)(2n_{m',n'}-1)$.
As the hyperhoneycomb lattice is bipartite, we term black ($b$) and white ($w$) sites so that the smaller-(larger-)$n$ site corresponds to the white (black) site on each $x$ bond as shown in Fig.~\ref{fig_hyperhoneycomb}(a).
Hence, the Hamiltonian in Eq.~(\ref{eq:5}) is rewritten as~\cite{Chen2007,Feng2007,Chen2008}
\begin{align}
{\cal H}&=J_x\sum_{x\, \textrm{bonds}}(a_w-a_w^\dagger)(a_b+a_b^\dagger)-J_y\sum_{y\, \textrm{bonds}}(a_b+a_b^\dagger)(a_w-a_w^\dagger)
-J_z\sum_{z\, \textrm{bonds}}(2n_b-1)(2n_w-1)\nonumber\\
&=iJ_x\sum_{x\, \textrm{bonds}}c_w c_b-iJ_y\sum_{y\, \textrm{bonds}}c_b c_w-iJ_z\sum_{z\, \textrm{bonds}} \eta_r c_b c_w.
 \label{eq:1}
\end{align}
In the second line of Eq.~(\ref{eq:1}), we introduced Majorana fermion operators, $c$ and $\bar{c}$, from the spinless fermion operators, $a$ and $a^\dagger$, as $c_w=(a_w-a_w^\dagger)/i$, $\bar{c}_w=a_w+a_w^\dagger$, $c_b=a_b+a_b^\dagger$, and $\bar{c}_b=(a_b-a_b^\dagger)/i$. 
In addition, we introduced $Z_2$ operators as $\eta_r=i \bar{c}_b \bar{c}_w$ on each $z$ bond $r$~\cite{Feng2007}. 
Since all the $Z_2$ operators commute with the Hamiltonian given by Eq.~(\ref{eq:1}), the eigenstates of the system are characterized by the set of eigenvalues $\eta_r=\pm 1$.
Note that we take open boundary conditions along the chains for avoiding a subtle boundary problem intrinsic to the Jordan-Wigner transformation.

The Hamiltonian in Eq.~(\ref{eq:1}) is a free Majorana fermion system coupled with the $Z_2$ variables $\{\eta_r\}$ on the $z$ bonds. 
The partition function of the system described by the Hamiltonian in Eq.~(\ref{eq:1}) is given by 
$Z={\rm Tr}_{\{\eta_r\}}{\rm Tr}_{\{c_i\}}e^{-\beta {\cal H}}={\rm Tr}_{\{\eta_r\}}e^{-\beta F_{f}(\{\eta_r\})}$
(we set the Boltzmann constant $k_{\rm B}$=1), and $F_f(\{\eta_r\})$ is the free energy of the Majorana fermion system for a given configuration of $\{\eta_r\}$: $F_{f}(\{\eta_r\})=-T \ln {\rm Tr}_{\{c_i\}}e^{-\beta {\cal H}(\{\eta_r\})}$.
Since the Hamiltonian ${\cal H}(\{\eta_r\})$ is given in a quadratic form in terms of the Majorana fermion operators, it is easily diagonalized in the form of
\begin{align}
{\cal H}(\{\eta_r\})=\sum_\lambda^{N/2}\varepsilon_\lambda(\{\eta_r\})\left(f_\lambda^\dagger f_\lambda -\frac{1}{2}\right),
\end{align}
where $f_\lambda$ ($f_\lambda^\dagger$) is the annihilation (creation) operator of a spinless fermion and $N$ is the number of lattice sites.
We perform the Markov-chain MC simulation for the classical local variables $\eta_r=\pm 1$ so as to reproduce the Boltzmann distribution of $e^{-\beta F_{f}(\{\eta_r\})}$.

We performed the replica exchange MC simulations for avoiding the freezing of MC sampling at low $T$, on the $L=4$, $5$, and $6$ clusters where $N=4L^3$~\cite{Hukushima1996}. 
We impose open boundary conditions for the $\bm{a}$ and $\bm{b}$ directions as mentioned above, and a periodic boundary condition for the $\bm{c}$ direction [see Fig.~\ref{fig_hyperhoneycomb}(a)]. 
We prepared 16 replicas and performed the single-flip update in the simulation for each replica.
We spent 40,000 (16,000) MC steps for measurement and 10,000 (1,000) MC steps for thermalization in the $L=4$ and $5$ clusters ($L=6$ cluster).

\section{Results}

\begin{figure}[t]
\begin{center}
\includegraphics[width=0.9\columnwidth,clip]{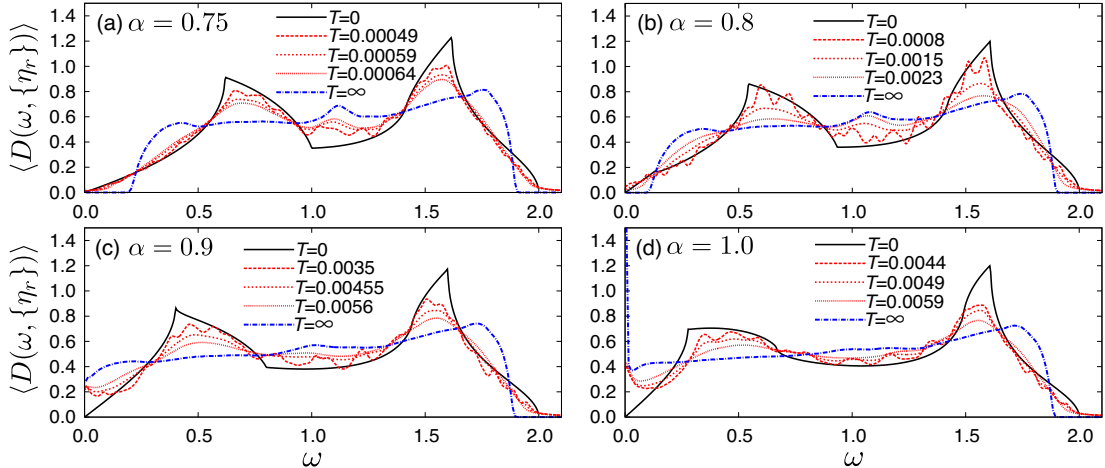}
\caption{
The DOS of Majorana fermions in the gapless region in Fig.~\ref{fig_hyperhoneycomb}(b): 
(a) $\alpha=0.75$, (b) $\alpha=0.8$, (c) $\alpha=0.9$, and (d) $\alpha=1.0$.
Except for the results at $T=0$ and $T=\infty$, the DOS are calculated by the MC simulation in $L=6$ clusters, where the smearing factor $\delta$ defined in Eq.~(\ref{eq:3}) is chosen to be 0.02.  The temperatures are taken in the vicinity of $T_c$ (see also Fig.~\ref{fig_lowene}).
}
\label{fig_dos}
\end{center}
\end{figure}

\begin{figure}[t]
\begin{center}
\includegraphics[width=0.9\columnwidth,clip]{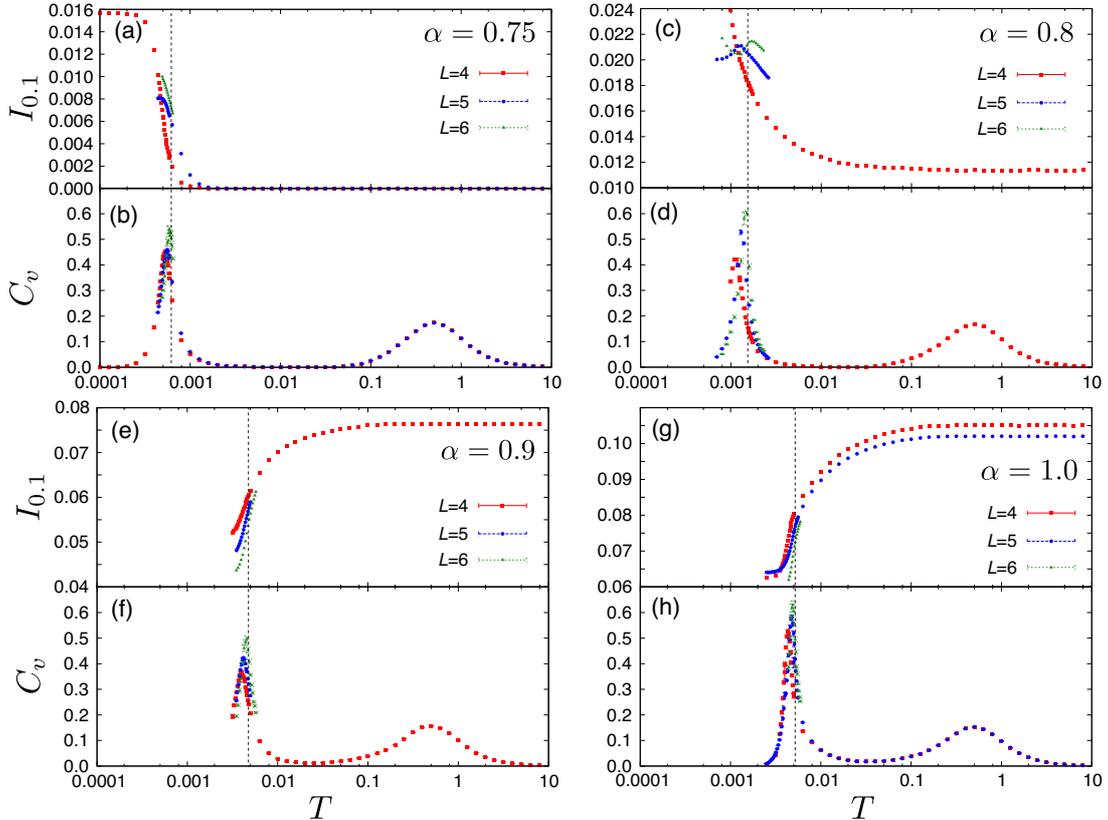}
\caption{
$T$ dependences of the low-energy weight of the DOS obtained by the integrations of the DOS in the range of $\omega=[0.0:0.1]$: 
(a) $\alpha=0.75$, (c) $\alpha=0.8$, (e) $\alpha=0.9$, and (g) $\alpha=1.0$.
The corresponding $T$ dependences of the specific heat $C_v$ are shown in (b), (d), (f), and (h). 
The vertical dotted line in each figure indicates $T_c$.
}
\label{fig_lowene}
\end{center}
\end{figure}

Before going into the MC results at finite $T$, let us first discuss the behavior of the DOS of Majorana fermions at zero $T$ and in the high-$T$ limit. 
The DOS is defined by
\begin{align}
 D(\omega,\{\eta_r\})=\frac{2}{N}\sum_\lambda \delta\left(\omega - \varepsilon_\lambda(\{\eta_r\})\right)=-\frac{1}{\pi}\frac{2}{N}\sum_\lambda{\rm Im}\frac{1}{\omega - \varepsilon_\lambda(\{\eta_r\}) + i\delta}\Bigg|_{\delta\rightarrow +0}.\label{eq:3}
\end{align}
The results at $T=0$ are easily obtained by performing the Fourier transformation for the Majorana fermions to diagonalize the Hamiltonian given in Eq.~(\ref{eq:1}), as the ground state is given by a uniform configuration of the $Z_2$ variables with all $\eta_r=+1$. 
On the other hand, the high-$T$ limit is given by random configurations of $\eta_r$: the DOS at $T=\infty$ is obtained by a simple average of Eq.~(\ref{eq:3}) over random configurations of $\{\eta_r\}$.  
The results at $T=0$ and $T=\infty$ are shown in Fig.~\ref{fig_dos} while changing the anisotropy parameter $\alpha$. 
The $T=0$ results are obtained by replacing the integrals by the sum over grid points of $300\times 300$ in the Brillouin zone, while the $T=\infty$ results are calculated for $L=12$ clusters with taking 2,400 random configurations. The parameter $\alpha$ is defined so as to satisfy $J_x=J_y=\alpha/3$ and $J_z=1-2\alpha/3$ 
[see Fig.~\ref{fig_hyperhoneycomb}(b)]; 
hence, the results in Fig.~\ref{fig_dos} are in the region where the ground state is gapless ($\alpha=0.75$ is critical).
Indeed, at $T=0$, the low-energy DOS is proportional to the excitation energy $\omega$ as shown in Fig.~\ref{fig_dos}, reflecting the Dirac-type semimetallic band structure. 
On the other hand, the DOS at $T=\infty$ shows contrasting behavior depending on the values of $\alpha$: an excitation gap opens for $\alpha\lesssim 0.8$, whereas the DOS becomes metallic with nonzero values at $\omega=0$ for $\alpha\gtrsim 0.9$.
 There is a boundary at $0.8 < \alpha_{c}(T=\infty) < 0.9$ between the gapped and gapless behavior in the high-$T$ limit. 
 This critical value of $\alpha$ is clearly larger than that at $T=0$, $\alpha_c(T=0)=0.75$. 
These results indicate that the Majorana fermion gap opens with increasing $T$ in the region of $\alpha_{c}(T=0)<\alpha<\alpha_{c}(T=\infty)$.

In order to clarify how the DOS evolves and the gap opens as $T$ increases, we calculate the thermal average of the DOS in Eq.~(\ref{eq:3}) by the MC simulation introduced in Sec.~\ref{sec:method}. 
The results are shown in Fig.~\ref{fig_dos} together with those at $T=0$ and $T=\infty$.
We here show the data in the vicinity of the critical temperatures $T_c$, which are estimated by the peak temperatures of the specific heat shown in Fig.~\ref{fig_lowene}~\cite{Nasu2014p}.
We can see that the low-energy part of the DOS changes rapidly near $T_c$, and develops a gap (fills a semimetallic dip) for $\alpha \lesssim (\gtrsim) \, \alpha_c(T=\infty)$.

To quantify the $T$ dependence of the low-energy DOS, we introduce the integral $I_\Omega$ of the low-energy part of the DOS defined by
$I_\Omega=\int_0^{\Omega} \langle D(\omega,\{ \eta_r \})\rangle d\omega$ 
(the bracket denotes the thermal average). 
Figure~\ref{fig_lowene} summarizes the values of $I_{\Omega=0.1}$, along with the specific heat $C_v$. 
The results clearly show that the DOS rapidly changes near $T_c$, where $C_v$ exhibits a sharp peak at $T_c$.
As shown in Figs.~\ref{fig_lowene}(a) and \ref{fig_lowene}(c), the low-energy weight of the DOS, $I_{0.1}$, rapidly decreases near $T_c$ as $T$ increases, reflecting the opening of the gap in the spectra for $\alpha \lesssim \alpha_c(T=\infty)$ at high $T$.
On the other hand, in the case of $\alpha>\alpha_c(T=\infty)$, $I_{0.1}$ rapidly increases in the vicinity of $T_c$, corresponding to the filling up the semimetallic dip, as shown in Figs.~\ref{fig_lowene}(e) and \ref{fig_lowene}(g). 
Thus, the low-energy Majorana fermion states are significantly modified by the phase transition from the low-$T$ QSL to high-$T$ paramagnet. 
As shown in our previous study~\cite{Nasu2014p}, the $Z_2$ fields are rapidly disordered near $T_c$. 
Hence, our results indicate that the thermal fluctuations affect the low-energy Majorana states through the $Z_2$ variables. 
Interestingly, the effect appears in a contrasting way below and above the boundary $\alpha_c(T=\infty)$, which is different from the quantum critical point $\alpha_c(T=0)=0.75$.

\section{Summary}
In summary, we have investigated the temperature variation of the Majorana fermion state in the 3D Kitaev model on the hyperhoneycomb lattice by using the Monte Carlo simulation. 
We found that the density of states of Majorana fermions evolves in a characteristic way in the gapless quantum spin liquid region. 
There is a clear boundary for the finite-temperature behavior: the Majorana fermion state develops an excitation gap in the region closer to the ground state phase boundary to the gapped region, whereas it is filled up to be metallic in the other region.
We showed that the evolution appears in the vicinity of the critical temperature for the spin liquid formation. 
Our results indicate that the low-energy spectra of Majorana fermions are significantly affected by thermal fluctuations in the $Z_2$ variables.

\ack
J.N. is supported by the Japan Society for the Promotion of Science through a research fellowship for young scientists.
This work is supported by Grant-in-Aid for Scientific Research (No. 24340076,
26400339, and 24740221), the Strategic Programs for Innovative Research (SPIRE), MEXT, and the Computational Materials Science Initiative (CMSI), Japan.
Parts of the numerical calculations are performed in the supercomputing systems in ISSP, the University of Tokyo.

\smallskip

\end{document}